\documentclass[12pt,preprint]{aastex}

\begin{document}

\newcommand{\escs}{erg\,s$^{-1}$cm$^{-2}$sr$^{-1}$}
\newcommand{\cmt}{cm$^{-3}$}
\newcommand{\cmd}{cm$^{-2}$}
\newcommand{\h}{H$_2$}
\newcommand{\kms}{km\,s$^{-1}$}
\newcommand{\um}{$\mu$m}
\newcommand{\lam}{$\lambda$}

\title{Spitzer spectral line mapping of protostellar outflows: II H$_2$ emission in L1157}

\author{Brunella Nisini\altaffilmark{1}, Teresa Giannini\altaffilmark{1}, 
David A. Neufeld\altaffilmark{2}, Yuan Yuan\altaffilmark{2}, 
Simone Antoniucci\altaffilmark{1},
Edwin A. Bergin\altaffilmark{3}, 
Gary J. Melnick\altaffilmark{4}}

\altaffiltext{1}{INAF - Osservatorio Astronomico di Roma Via di Frascati 33,
00040, Monteporzio Catone, Italy}

\altaffiltext{2}{Department of Physics and Astronomy, John Hopkins University,
  3400 North Charles Street, Baltimore, MD 21218}
   
\altaffiltext{3}{Department of Astronomy, University of Michigan, 500 Church
  Street, Ann Arbor, MI 48109-1042}
  
\altaffiltext{4}{Harvard-Smithsonian Center for Astrophysics, 60 Garden
  Street, Cambridge, MA 02138}
  

\begin{abstract}

We present an analysis of Spitzer-IRS spectroscopic maps of the L1157 protostellar 
outflow in the \h\ pure-rotational lines from S(0) to S(7). 
The aim of this work is to derive the physical conditions pertaining to the
warm molecular gas and study their variations within the flow.
The mid-IR \h\ emission follows the morphology of the precessing flow, with
peaks correlated with individual CO clumps and \h\, 2.12\um\, ro-vibrational emission. 
More diffuse emission delineating the CO cavities is detected only in
the low-laying transitions, with $J_{lower} \le$ 2.
The \h\, line images have been used to construct 2D maps of $N(H_2)$, \h\ ortho-to-para ratio
and temperature spectral index $\beta$, in the assumption of a gas temperature stratification 
where the \h\ column density varies as $T^{-\beta}$. 
Variations of these parameters are observed along the flow. 
In particular, the ortho-to-para ratio ranges from $\sim$0.6 to 2.8, highlighting the presence of regions 
subject to recent shocks where the ortho-to-para ratio has not had time yet to reach
the equilibrium value. 
Near-IR spectroscopic data on ro-vibrational \h\ emission have been combined with the 
mid-IR data and used to derive additional shock parameters in the brightest blue- and red-shifted 
emission knots. A high 
abundance of atomic hydrogen (H/\h $\sim$ 0.1-0.3) is implied by the 
observed \h\ column densities, assuming n(\h) values as derived by independent 
SiO observations. The presence of a high fraction of atomic hydrogen, 
indicates that a partially-dissociative
shock component should be considered for the \h\ excitation in these localized regions.  
However, planar shock models, either of C- or J-type, are
not able to consistently reproduce all the physical parameters derived from our analysis 
of the \h\ emission. 
Globally, \h\ emission contributes to about 50\% of the total shock radiated energy in the L1157 outflow. We find that the momentum flux through the shocks derived from the radiated luminosity is
comparable to the thrust of the associated molecular outflow, supporting the scenario
where this latter is driven by the shock working surface. 
\end{abstract}



\section{Introduction}
The interaction of protostellar outflows with the ambient molecular cloud occurs through radiative
shocks that compress and heat the gas, which in turn cools down through line emission at
different wavelengths. In the dense medium where the
still very embedded protostars (the so called class 0 sources) are located, shocks are primarily non-dissociative, and hence the cooling is mainly through emission from 
abundant molecules. Molecular hydrogen is by far the most abundant species in these environments, 
and although \h\, emits only through quadrupole transitions with low radiative rates,
it represents the main gas coolant in  flows from young protostars. \h\, shocked emission in outflows has been widely studied in the past mainly through its
ro-vibrational emission in the near-IR (e.g. Eisl{\"o}ffel et al. 2000; Giannini et al. 2004; Caratti o Garatti et al. 2006) that traces the dense
gas at T$\sim$2000-4000 K. Most of the thermal energy associated with the shocks is however radiated away through
the emission of \h\ rotational transitions of the ground state vibrational level at $\lambda \le 28\mu$m
(e.g. Kaufman \& Neufeld 1996). 
Mid-IR \h\, lines are easily excited at low densities and temperatures between 300 and 1500 K: therefore they
are very good tracers of the molecular shocks associated with the acceleration of ambient gas by
matter ejection from the protostar. Given the low excitation temperature, 
they can also probe regions where \h\, has not yet reached the ortho-to-para equilibrium, thus 
giving information on the thermal history of the shocked gas (Neufeld et al. 1998; Wilgenbus et al. 2000).
In addition, given the different excitation temperature and critical densities of the v=0--0 and v$\ge$ 1 \h\ lines,
the combination of mid-IR with near-IR observations is a very powerful tool to constrain 
the global physical structure and the shock conditions giving rise to the observed emission.

The study of the 0--0 rotational emission in outflows started in some detail with the 
\emph{Infrared Space Observatory}. Thanks to the observations performed with the SWS and ISOCAM instruments,
the shock conditions, the ortho-para ratio and the global \h\, cooling  have been derived in a handful of 
flows (e.g. Neufeld et al. 1998; Nisini et al. 2000; Molinari et al. 2000;  Lefloch et al. 2003). More recently, the \emph{Infrared Spectrometer} (Houck et al. 2004) on board
\emph{Spitzer}, with its enhanced spatial resolution and sensitivity with respect to the ISO spectrometers, 
has been used to obtain detailed images of the \h\, rotational emission, from S(0) to S(7), of several
outflows, from which maps of  important physical parameters, such as temperature, column density and o/p
ratio, have been constructed (Neufeld et al. 2006; Maret et al. 2009;  Dionatos et al. 2010). 
In this framework, Neufeld et al. (2009, hereafter N09)  have recently presented IRS spectroscopic maps observations of five young 
protostellar outflows at wavelengths between 5.2 and 37\um, and discussed their averaged physical properties 
and overall energetics. In all the flows, the \h\, S(0)-S(7) emission has been detected and contributes to more than 95\%  of the 
total line luminosity in the 5.2-37\um\, range, while atomic emission, in the form  of FeII and SI fine structure lines, accounts for only the remaining $\sim$5\%. 

In the present paper, we will analyse the \h\, line maps obtained by N09 towards the L1157 outflow,
with the aim of deriving the main physical conditions pertaining to the molecular gas and their variations 
within the flow. This in turn will give information on the thermal history of the flow and on how 
energy is progressively transferred from the primary ejection event to the slow moving ambient gas. 

For this first detailed analysis, L1157 has been chosen among the sources observed by N09 given its uniqueness as a very active and well studied flow at different wavelengths.   
More than 20 different chemical species have been indeed 
detected in the shocked spots of this object (Bachiller \& Perez Gutierrez 1997; Benedettini et al. 2007), some of them for the first time in outflows (e.g. HNCO, Rodr{\'{\i}}guez-Fern{\'a}ndez et al. 2010, and complex organic molecules, Arce et al. 2008, Codella et al. 2009)
, testifying for a rich shock induced chemistry. Warm \h\ shocked emission in L1157
is also evidenced through near-IR maps (e.g. Davis \& Eisl\"offel, 1995) and Spitzer-IRAC
images (Looney et al. 2007). The L1157
outflow has been also recently investigated with the \textit{Herschel Space Observatory}, showing
to be very strong also at far-IR wavelengths (Codella et al. 2010, Lefloch et al. 2010,
Nisini et al. 2010).

The L1157 outflow extends about 0.7 pc in length. Its distance is uncertain and has been estimated between 250 and 440 pc. Here we will adopt D=440 pc for an easier comparison with other works.
The outflow is driven by a highly embedded, low mass class 0 source (L1157-mm or IRAS20386+6751) having $L_{bol} \sim $ 8.3 $L_\odot$ (Froebrich 2005). 
It is a very nice example of an outflow driven by a precessing and pulsed jet, possessing an S-shaped structure and different cavities, whose morphology has been reproduced assuming that the outflow
is inclined by $\sim$80$^\circ$ to the line of sight and the axis of the underlying jet precesses
on a cone of 6$^\circ$ opening angle (Gueth et al. 1996). The episodic mass ejection events are
evidenced by the presence, along the flow, of individual clumps that are symmetrically displaced 
with respect to the central source.
It is therefore a very interesting target for a  study of the physical conditions pertaining to these active regions through an \h\ excitation analysis.

The paper is organized as follow: the observations and the main results are summarized in \S 2. In \S 3 
we describe the analysis performed on the \h\ images to derive maps of temperature, column density and 
ortho-to-para ratio. A more detailed NLTE analysis on individual emission peaks is also presented here, where the 
Spitzer data are combined with near-IR data to further constrain the excitation conditions.  The implications of these results for the shock conditions along the L1157  flow are discussed in \S 4, together with an analysis of the global
energy budget in the flow.  A brief summary follows in $\S 5$.

\section{Observations and results \label{analysis}}

Observations of the L1157 outflow were obtained in November 2007 with the IRS instrument, during Cycle 4 of the Spitzer mission . 
The full IRS spectral range (5.2-36.5$\mu$m) was observed with the Long-High (LH), Short-High (SH) (R $\sim$ 600) 
and Short-Low (SL) (R between 64 and 128) modules. 
The L1157 outflow region was covered through 5  individual IRS maps of $\sim$ 1\arcmin x1\arcmin\,  of size each, arranged along the outflow axis. Each map was obtained by stepping the IRS slit by half of its width in the direction perpendicular
to the slit length. For the SH and LH modules the slit was stepped also parallel to its axis by 4/5 (SL) and 1/5 (LH)  of its length. 
Details on the data reduction that generated the individual line maps from the IRS scans are given in N09. The final maps have been resampled to a grid of 2\arcsec\, spacing allowing a pixel by pixel comparison of maps obtained with the different IRS modules.
Maps of the brightest detected lines as well as the full spectrum in a representative position are shown in  Fig.\,7 and Fig.\,12 of N09. As regards to \h, all the pure rotational lines of the first vibrational levels, from S(0) to S(7),
 are detected at various intensity along the flow. Here we report, in Tab. \ref{fluxes},  the \h\, brightness measured in a 20\arcsec\, FWHM Gaussian aperture towards different positions.

 Fig\,1  shows the L1157 maps of the S(1) and S(2) lines while Fig.\,2 displays the S(5) line with superimposed contours of the CO 2--1 emission from Bachiller et al. 2001.
 In the same figure, a map of the 2.12 $\mu$m 1--0 S(1) line is also presented. The morphology of the 0--0 S(5) and 1--0 S(1) is very similar, with  peaks of mid-IR emission 
located at the near-IR knots from A to D, as identified by Davis \& Eisl\"offel (1995). 
 When compared with the CO map, the mid-IR \h\, emission appears to follow the curved chain of clumps (labelled as B0-B1-B2 and R0-R1-R, for the blue-shifted and red-shifted lobes, respectively) that also correspond to peaks of SiO emission, as resolved in interferometric observation by Gueth et al. (1998) and Zhang et al. (2000). 
 The L1157 outflow  morphology has been suggested to delineate a precessing flow (Gueth et al. 1996), where the \h\, and SiO peak emission knots follow  the location of the actual working surface of the precessing jet and are thus associated with the youngest ejection episodes. 
Diffuse \h\ emission is also detected in the S(1)-S(2) maps, that delineates the wall of a cavity that connects the central source with both
the R0 and B0 clumps. Such a cavity has been recognized in the CO 1-0 interferometric maps and it is likely created by the propagation of large bow-shocks.
The S(1)-S(2) maps of Fig.\,1 show  extended emission of \h\,  also in the SE direction (i.e. where the B2 clump is located) and in the eastern edge of the northern lobe, that also follow  quite closely the CO morphology: these regions at lower excitation might trace additional cavities created by an older ejection episode of the precessing jet.

\section{ H$_2$ Analysis }
\subsection{LTE 2D analysis of the rotational lines: maps of averaged parameters }
\label{sec:maps}
We have used the  \h\, line maps to obtain the 2D distribution of basic \h\, physical parameters, through the analysis 
of the rotational diagrams in each individual pixel. As described in N09, who analysed the global 
\h\, excitation conditions in L1157, the distribution of upper level column densities of the S(0)-S(7) lines as a function of their 
excitation energy, does not follow a straight line, indicating that a  temperature stratification in the 
observed medium exists. The exact form of this temperature stratification depends on the type of shock
the \h\, lines are tracing, as they probe the post-shock regions where the gas cools from $\sim  $ 1000 K 
to $ \sim $ 200 K. 

The simplest way to parametrize the post-shock temperature stratification
is to assume a power-law distribution: this approach was applied by 
Neufeld \& Yuan (2008), which also show that this type of distribution is expected  
in gas ahead of unresolved bow-shocks. On this basis, and following also N09, we fit the observations 
assuming a slab of gas where the \h\, column density in each layer at a given $T$ varies as 
$ dN \propto T^{-\beta}dT$.

This law is integrated, to find the total column density, between a minimum ($T_{min}  $) and a maximum ($T_{max}$) temperatures . For our calculations 
we have kept $T_{max}$ fixed at 4000 K, since gas at temperatures larger than this value is not expected to contribute to the emission of the observed pure rotational lines. $T_{min} $ was instead assumed to be equal, in each position, to the minimum temperature probed by the observed lines. This $T_{min} $ is taken as the excitation temperature giving
rise to the observed ratio of the S(0) and S(1) column densities, assuming a Boltzman
distribution. 
The $T_{min}  $ value ranges between $\sim$150 and 400 K.

We found that the approach of a variable  $T_{min}$
produces always fits with a better $\chi^2$ than assuming a fixed low value in all positions.
We also assume the gas is in LTE conditions. Critical densities of rotational lines from S(0) to S(7)
range between 4.9 \cmt\, (S(0)) and 4.4$\times$10$^5$\cmt(S(7)) at T=1000 K assuming only \h\, collisions
(Le Bourlot et~al. 1999): critical densities decrease if collisions with H and He are not negligible. 
Deviations from LTE can be therefore expected only for the high-$J$ S(6) and S(7) transitions: the S/N of these 
transitions in the individual pixels is however not high enough to disentangle, in the rotational diagrams, NLTE 
effects from the effects caused by the variations of the other considered parameters. In particular, as also
discussed in  N09, there is a certain degree of degeneracy in the density and the $\beta$ parameter 
of the temperature power law in a NLTE treatment that we are not able to remove in the analysis of the
individual pixels. This issue will be further discussed in \S \ref{sec:NLTE}. 
An additional parameter of our fit is the ortho-to-para ratio (OPR) value. It is indeed recognized that the 0--0 \h\, lines are
often far from being in ortho-to-para equilibrium, an effect that in a rotational diagram is evidenced by
a characteristic zigzag behavior in which column densities of lines with even-$J$ lie systematically above those of odd-$J$ lines. In order not to introduce too many parameters, we assume a single OPR value as a free parameter for the
fit. In reality, the OPR value is temperature dependent (e.g. N09 and  \S \ref{sec:NLTE}), and therefore
the high-$J$ lines  might present an OPR value closer to equilibrium then the low-$J$ transitions. 
Our fit gives therefore only a value averaged over the temperature range probed by the lines considered (i.e. $\sim$200-1500 K). 

In summary, we have varied only three parameters, namely the total \h\, column density N(\h), the OPR, and the temperature power law index $\beta$,  in order to obtain the best model fit through a $\chi^2$ minimization procedure and assuming a 20$\%$ flux uncertainty for all the lines. 
The fit was performed only in those pixels where at least four lines with an S/N larger
than 3 have been detected.  
Before performing the fit, the line column densities were corrected for extinction assuming $A_v$ = 2 (Caratti o Garatti et al. 2006) and 
adopting the Rieke \& Lebofsky (1985) extinction law. At the considered wavelengths,
variations of $A_v$ of the order of 1-2mag do not affect any of the derived results.

Figure \ref{fig:ncol} shows the derived map of the \h\, column density, while in Fig. \ref{fig:b_tmin} and \ref{fig:op_tcold}  
maps of the OPR and  $\beta$ are displayed together with temperature maps relative to the ''cold'' and ''warm'' components ($T_{cold}$ and $T_{warm}$),
i.e. the temperature derived from linearly fitting the S(0)-S(1)-S(2) and S(5)-S(6)-S(7) lines, once corrected for the derived OPR value. 
In Fig. 6 we also show the individual excitation diagrams for selected positions along the flow, obtained from intensities
measured in a 20$"$ FWHM Gaussian aperture centered towards emission peaks (Tab. 1). Values of the
fitted parameters in these positions are reported in Tab. \ref{param}. In addition to $T_{cold}$ and $T_{warm}$,
we give in this table also the values of the average temperature in each knot, derived through a
linear fit through all the \h\ lines ($T_{med}$).

The maps show significant variations in the inferred parameters along the outflow. 
The \h\ column density ranges between 5$\times$10$ ^{19} $ and 
3$\times$10$ ^{20} $\cmt. The region at the highest column density is located towards the B1 molecular bullet (see Fig. \ref{fig:h2co})\footnote{Fig. \ref{fig:h2co} shows that the molecular clumps B1, R0 and
R coincide in position with the NIR \h\ knots A, C and D. In the paper, both nomenclature
will be used specifying if we refer to the NIR or mm condensations}. 
This is consistent with the higher column density of CO found in B1 with respect 
to other positions in the blue lobe (Bachiller \& Perez Gutierrez 1997), and might suggest that this is a zone where the outflowing gas is compressed due 
to the impact with a region of higher density (Nisini et al. 2007). Towards the NW, red-shifted 
outflow, the column density has a more uniform distribution, with a plateau at $ \sim 10 ^{20} $ \cmd\ that follows the \h\ intensity distribution. 
The N(\h) decreases at the apex of the red-shifted outflow, with a value slightly below $ 10 ^{20} $ \cmd\ at the position
of the D near-IR knot.

$T_{cold}$ ranges between $ \sim $ 250 and 550 K. The highest values are found at the tip of the northern outflow lobe, while local maxima corresponds to the positions of line intensity peaks.
$T_{warm}$ ranges between  $ \sim $ 1000 and 1500 K. In this case the highest values are in the southern lobe, at the position of the A NIR knot. 
As a general trend, the $T_{warm}$ value decreases going from the southern to the northern peaks of emission, with
the minimum value at the position of the D NIR knot. 

$\beta$ values range between $ \sim $4-4.5 in the blue-shifted lobe while it is larger in the red-shifted
lobe, with maximum values of $ \sim $ 5.5 at the tip of the flow. 
Due to the degeneracy between $\beta$ and density discussed in the previous section, these
values can be considered as upper limits because of our assumption of LTE conditions. 
Neufeld \& Yuan (2008) have discussed the $\beta$ index expectations in bow-shock excitation.
A $\beta$ index of $ \sim $ 3.8 is expected in paraboloid bow shocks having a velocity at the bow apex
high enough to dissociate H$_2$,  in which case the temperature distribution
extends to the maximum allowed temperature. Slower shocks that are not able to attain the maximum
temperature, produce steeper temperature distributions, i.e. with values of $ \beta $ greater than 3.8.
This is consistent with our findings: low values of $ \beta $ (of the order of 4) are found in the blue-shifted
lobe: here, evidence of \h\ dissociation is given by the detection
of atomic lines (i.e. [SiII], [FeII] and [SI]) in the IRS spectrum. In the red-shifted lobe, where values of $\beta$ larger than 4 are derived in the LTE
assumption, no atomic emission is detected and the $T_{warm}$ values
are lower than those measured in the blue-shifted lobe, indicating a maximum temperature lower than in the blue flow .

The OPR varies significantly along the outflow, spanning from $ \sim $ 0.6 to 2.8. Hence it is always below the equilibrium value of 3.  Although a one-to-one correlation between temperature and OPR cannot be discerned, some trends can be
inferred from inspection of Fig. \ref{fig:op_tcold}. In general the OPR minima are observed in plateau regions between two consecutive
intensity peaks, where also the cold temperature has its lowest values. At variance with this trend, the emission
filaments delineating the outflow cavity within $\pm$ 20$ \arcsec  $ from the mm source, where the cold 
temperature reaches a minimum value of $ \sim $ 250 K,   show rather high
values of the OPR, $ \sim $2.4-2.8. This might suggests that this region has experienced an older 
shock event that has raised the OPR, though not to the equilibrium value, and where the gas
had time to cool at a temperature close to the pre-shock gas temperature. 
On the other hand, at the apex of both  the blue- and red-shifted lobes, where the cold temperatures 
are relatively high (i.e. 500-550 K), the OPR is rather low, $ 1.5-2.0 $. Evidence of regions of low OPR and high 
temperatures at the outflow tips has been already given in other flows (Neufeld et al. 2006; Maret et al. 2009). It has been suggested that these represent zones subject to recent shocks where the OPR has not had time yet to reach the
equilibrium value.

\subsection{ NLTE analysis: constraints on H and H$_2$ particle density}
\label{sec:NLTE}

Additional constraints on the physical conditions responsible for the \h\, excitation, are provided by combining
the emission of the mid-IR \h\ pure-rotational lines from the ground vibrational level with the emission from near-IR ro-vibrational lines.
It can be seen from Fig. \ref{fig:h2co} that the 2.12\um\, emission follows quite closely the
emission of the 0--0 lines at higher $J$. In addition to the 2.12\um\, data presented in Fig. \ref{fig:h2co}, we have also considered
the NIR long-slit spectra obtained on the A and C NIR knots by Caratti o Garatti (2006). These knots, at the spatial
resolution of the NIR observations (i.e. $\sim$0.8\arcsec), are separated in several different sub-structures that have been individually investigated with the long-slit spectroscopic observations. For our analysis we have considered the
data obtained on the brightest of the sub-structures,  that coincide, in position, 
with peaks of the 1--0 S(1)  line.

In order to inter-calibrate in flux the Spitzer data with these NIR long-slit data, obtained  with a slit-width of 0.5 arcsec, 
we have proceeded as follows: we first convolved the 
2.12\um\, image at the resolution of the Spitzer images and than performed photometry on the
A and C peaks positions with a 20\arcsec\, diameter Gaussian aperture, i.e. with the same
aperture adopted for the brightness given in Tab. \ref{fluxes}. We have then scaled the fluxes of the individual lines given in 
Caratti o Garatti (2006) in order to match the 2.12\um\ flux gathered in the slit with that measured by the
image photometry. In doing this, we assumed that the average excitation conditions within the 20$\arcsec$ aperture are
not very different from those of the A-C peaks. This assumption is 
observationally supported by the fact that the ratios of different \h\ NIR lines 
do not change significantly (i.e. less than 20\%) in the A-C knot substructures separately
investigated in Caratti o Garatti (2006).
We have considered only those lines detected with S/N larger 
than 5; in practice this means considering lines from the first four and three vibrational levels for knots A and C, respectively. 
The excitation diagrams obtained by combining the Spitzer and NIR data for these two knots are displayed in Fig.\ref{fig:ACfit}.

In order to model together the 0--0 lines and the near-IR ro-vibrational lines, we have implemented
 two  modifications to the approach adopted previously. First of all, the NIR lines probe gas
at temperatures higher than the pure rotational lines, of few thousands of K, at which values it is expected
that the OPR has already reached equilibrium. Thus, 
the ortho-para conversion time as a function of temperature needs to be included in the
fitting procedure, since lines excited at different temperatures have different OPRs.
We  have here adopted the approach of N09 and used an analytical expression for the OPR as a function of the temperature, considering a gas that had an initial value of the ortho-to-para ratio OPR$_0  $ and has been heated to a temperature T for a time $  \tau$. Assuming that the para-to-ortho conversion occurs through reactive collisions with atomic hydrogen, we have:

\begin{equation}
\frac{OPR(\tau)}{1+OPR(\tau)}  = \frac{OPR_0}{1+OPR_0}\,e^{-n(H)k\tau} + {\frac{OPR_{LTE}}{1+OPR_{LTE}}}\,\left(  1 - e^{-n(H)k\tau}\right) 
\end{equation}

In this expression, n(H) is the number density of atomic hydrogen and OPR$_{LTE}$ is the ortho-to-para ratio equlibrium value.
The parameter $ k $ is given by the sum of the rates coefficients for para-to-ortho conversion ($ k_{po} $), estimated
as 8$ \times $10$ ^{-11} $exp(-3900/T) cm$ ^{3} $\,s$ ^{-1} $, and for ortho-to-para conversion, $ k_{op} \sim k_{po}/3 $
(Schofield et al. 1967). Thus the dependence of the OPR on the temperature is implicitly given by the dependence on T of the
$ k $ coefficient. The inclusion of a function of OPR on $T$, introduces one additional parameter to our fit: while we have previously 
considered only the average OPR of the 0--0 lines, we will fit here the initial OPR$_0  $ value and the coefficient $ K = n(H)\tau $.

The second important change that we have introduced with respect to the previous fitting procedure, 
is to include a NLTE treatment of the \h\, level column densities. In fact, the critical densities
of the NIR lines are much higher than those of the pure rotational lines (see Le Bourlot et~al. 1999). For example, the 
$n _{crit} $ of the 1-0 S(1) 2.12\um\, line is 10$^7$\cmt\, assuming only collisions with \h\, and T=2000 K. 
Therefore the previously adopted LTE approximation might not be valid when combining lines from different 
vibrational levels.
This is illustrated in Fig.\ref{fig:plot_dens}, where we plot the results obtained by varying the \h\ density between 
10$ ^{3} $ and 10$ ^{7} $\cmt\, while keeping the other model parameters fixed . The observed column densities in the A
position are displayed for comparison. 
For the NLTE statistical equilibrium computation we have adopted the \h\, collisional rate coefficients given by Le Bourlot et~al. (1999) \footnote{The  rate coefficients for collisions with ortho- and para-H$_2$, HI and He, computed in Le Bourlot et~al. (1999) 
are available at the web-site: http://ccp7.dur.ac.uk/cooling$\_$by$\_$h2/} .
This figure demonstrates the sensitivity of the relative ratios  between 0--0 and 1--0 transitions to density variations.   For example,  in this particular case, the ratio N(\h)$_{0-0 S(7)}$/N(\h)$_{1-0 S(1)}$ is 64.6 at n(\h)=10$ ^{4} $ \cmt\, and 1.9 at n(\h) $ \gtrsim $ 10$ ^{7} $ \cmt. The figure also shows that the observational points display only a small misalignment in column densities between the 0--0 lines and the 1--0 lines, already indicating that the ro-vibrational lines are close to
LTE conditions at high density.

In Fig.\ref{fig:ACfit}, we show the final best-fit models for the combined mid- and near-IR column densities 
in the A and C positions. 
As anticipated, the derived n(\h) densities are large, of the order of 10$ ^{7} $ and 6$\times$10$ ^{6}$ \cmt,  for the A and C positions, respectively. The two positions indeed show very similar excitation conditions: only the column density is a factor of 3 smaller in knot C. 
Hence, we conclude that the lack of detection of rotational lines with v$ > $ 3 in knot C, in contrast to knot A (Caratti o Garatti et al. 2006), is due
purely to a smaller number of emitting molecules along the line of sight and not to different excitation conditions.

The derived \h\, densities are much higher than  previous estimates based on other tracers. Nisini et al. (2007)
derive a density of 4$\times$10$ ^{5} $ \cmt\, at the position of knot A from multi-line SiO observations, thus more than an order of magnitude smaller than those inferred from our analysis. The high-J CO lines observed along the blue-shifted lobe of L1157 by 
Hirano et al. (2001), indicate a density even smaller, of the order of 4$\times$10$ ^{4} $ \cmt. 
SiO is synthesized and excited in a post-shock cooling zone where the maximum compression is reached, therefore
it should trace post-shock regions at densities higher than \h (e.g. Gusdorf et al. 2008). 
One possibility at the origin of the discrepancy is our assumption of collisions with only \h\ molecules, and thus of a 
negligible abundance of H. This can be considered roughly true in the case of non-dissociative C-shocks, where H atoms are 
produced primarely in the chemical reactions that form H$_2$O from O and \h, with an abundance n(H)/n(\h+H) $ \sim  $ 10$^{-3}$
(e.g. Kaufmann \& Neufeld 1996). However, if the shock is partially dissociative, the abundance of H can increase considerably and 
collisions with atomic hydrogen cannot be neglected, in view of its  large efficiency in the \h\, excitation.
This situation cannot be excluded at least for the knot A, where atomic emission from [FeII] and [SI] has been 
detected in our Spitzer observations.
Since we cannot introduce the n(H) as an additional independent parameter of our fit, we have fixed n(\h) at the value
derived from SiO observations (4$\times$10$ ^{5} $ \cmt, Nisini et al. 2007) and varied the H/\h\, abundance ratio. 
The best fit is in this case obtained with a ratio H/\h=0.3: this indicates that our observational data are
consistent with previous \h\ density determinations only if a large fraction of the gas is in atomic form.

Turning back to the inferred OPR variations with temperature, our fit implies that the OPR in the cold gas component at T=300 K
is significantly below the equilibrium value, while the value of  
OPR=3 is reached in the hot gas at T=2000 K traced by the NIR lines. The parameter  $K=n(H)\tau$
is constrained to be $\sim$10$^6$ and 10$^7$ yr\,\cmt\, for knots A and C, respectively. 
We can also estimate the time needed for the gas to reach this distribution of OPR, from the limits on the
atomic hydrogen abundance previously discussed.
Our data implies a high value of the n(H) density: a minimum value of n(H) $\sim$0.6-1$\times$10$^4$\cmt, (for knots C and A, respectively) 
 is given if we assume H/\h$\sim$ 10$^{-3}$ (and thus the n(\h) $\sim$ 10$ ^{7} $\cmt, given
 by our fit), while a maximum value of $\sim$10$^5$\cmt,
is derived from the fit where n(\h) is kept equal to 4$\times$10$ ^{5} $ \cmt.
The high abundance of atomic H ensures that conversion of para- to
ortho-\h\, proceeds very rapidly: the fitted values of the K parameter indeed imply that the observed range of
OPR as function of temperature have been attained in a timescale between 100 and 1000 yrs for both the knots.

Finally, given the column density and particle density discussed above, we can estimate the \h\,
cooling length ($L \sim$ N(\h)/n(\h)). If we consider the case of n(\h) $\sim$ 10$ ^{7} $\cmt,
and negligible n(H), we have $L \sim$ 10$ ^{13} $ cm while a length of $\sim$ 10$ ^{15} $ cm is inferred
in the case of n(\h)$\sim$ 4$\times$10$^5$\cmt. 
All the parameters derived from the above analysis are summarised in Tab. \ref{shock} and they will be discussed
in the next section in the framework of different shock models.

\section{Discussion}

\subsection{Shock conditions giving rise to the \h\, emission}

The copious \h\, emission at low excitation observed along the L1157 outflow 
indicates that the interaction of the flow with the ambient medium occurs
prevalently through non-dissociative shocks. 
Both the Spitzer IRS maps of N09, and the NIR narrow band images of Caratti o Garatti et al. 2006, show that significant gas dissociation in L1157 occurs only at the A peak, where both mid-IR 
lines from [FeII], [SII] and [SI] and weak [FeII] at 1.64\um\ have been detected. 
Weaker [SI]\,25\um\ and [FeII]\,26\um\ emission have been also detected on the C spot, but
overall the atomic transitions give a negligible contribution to the total gas cooling,
as pointed out in N09.
These considerations suggest that most of the shocks along the outflow occur at speeds
 below $\sim$ 40\kms,  as this is the velocity limit above which \h\ is expected to be dissociated.
The knot A is the only one showing a clear bow-shock structure. Here the velocity at
the bow apex is probably high, causing \h\ dissociation and atomic line excitation,
consistent with the fitted temperature power law $\beta$ of $\sim$ 4, as discussed in \S 3.1,
while the bulk of the \h\, emission comes from shocks at lower velocities originating in
the bow wings.

Constraints on the shock velocity that gives rise to the molecular emission in L1157
 have been already given in previous works. The sub-mm SiO emission and
abundances, measured in different outflow spots, suggest shock velocities of the 
order of 20-30\kms\ (Nisini et al. 2007). The comparison of SiO and \h\ 
emission against detailed shock models performed by Gusdorf et al. (2008) confirm a similar
range of velocities in the NIR-A knot, although the authors could not find a unique 
shock model that well represents both the emissions. 

Cabrit et al. (1999) found that the column density of the mid-IR \h\, emission lines, 
from S(2) to S(8), observed by ISO-ISOCAM was consistent either with C-shocks having
velocities of $\sim$\,25\kms\, or with J-shocks at lower velocity, of the order
of $\sim$\,10\kms. Gusdorf et al. (2008), however, conclude
that stationary shock models, either of C- or J-type, are not able to reproduce the observed
rotational diagram on the NIR-A position, constructed combining ISOCAM data and 
NIR vibrational lines emission. A better fit was obtained 
by these authors considering non-stationary shock models, which have developed a magnetic precursor but which retain a J-type discontinuity (the so-called CJ shocks, Flower et al. 2003). 
Similar conclusions, but on a different outflow, have been reached by Giannini et al. 2006
who studied the \h\ mid- and near-IR emission in HH54: in general, stady-state C- and J-type 
shocks fail to reproduce simultaneously the column densities of both the ro-vibrational 
and the v=0, pure-rotational \h\ levels.  

A different way to look at the issue of the prevailing shock conditions in the observed
regions, is to compare the set of physical parameters that we have inferred from our analysis 
to those expected from different shocks. With this aim, we summarize in   
Tab. \ref{shock} the physical properties derived on the A and C \h\, knots. 
In addition to the parameters derived from the NLTE analysis 
 reported in Section \ref{analysis}, namely H$_2$ post-shock density, H/\h\ fraction, 
 initial OPR, cooling length and time, the table reports also the average values of 
OPR and rotational temperature, as they are measured from a simple linear fit
 of the rotational diagrams presented in Fig. \ref{fig:fit_nir}.

As mentioned in \S 3.2, the high fraction of atomic hydrogen inferred by our analysis rule out excitation in a pure
 C-shock. In fact, dissociation in C-shocks is always too low to have a 
H/H$_2$ ratio higher than 5$\times$10$^{-3}$, irrespective from the shock velocity and magnetic field strength 
(Kaufman \& Neufeld 1996; Wilgenbus et al. 2000).
C-shocks are not consistent with the derived parameters even if we consider the model fit with
the high \h\ post-shock density of the order of 10$^{7}$ \cmt and negligible atomic hydrogen: in this case we derive an emission length of 10$^{13}$ cm,
which is much lower than the cooling length expected in C-shocks, which, although 
decreasing with the pre-shock density, is never less than 10$^{15}$ cm (Neufeld et al. 2006) .

Stationary J-shock models better reproduce some of our derived parameters.
For example, in J-type shocks the fraction of hydrogen in the post-shocked gas
can reach the values of 0.1-0.3 we have inferred, provided that the shock velocity is larger
than $\sim$ 20 \kms. In general, a reasonable agreement with the inferred post-shock density and 
H/\h\ ratio is achieved with models having $v_s$=20-25 \kms and pre-shock densities of 10$^3$\cmt
(Wilgenbus et al. 2000). Such models predict a shock flow time of the order of 100 yr or less, which
is also in agreement with the value estimated in our analysis at least in knot A.
 In such models, however, the cooling length is an order of magnitude smaller than the 
inferred value of $ \sim $10$ ^{15} $ cm. In addition, the gas temperature remains high for most of the post-shocked region: the average rotational temperature of 
the v=0 vibrational level is predicted to be, according to the Wilgenbus et al. (2000)
grid of models, always about 1600 K or larger, as compared with the value of about 800-900 K inferred from observations. 
The consequence of the above inconsistencies is that J-type shocks tend to underestimate 
the column densities of the lowest \h\, rotational levels in L1157, an effect already pointed 
out by Gusdorf et al. (2008). 

As mentioned before, Gusdorf et al. (2008) conclude that the \h\ pure rotational emission in L1157 
is better fitted with a non-stationary C+J shock model with either $v_s$ between 20 and 25 \kms\ and pre-shock densities $n_H = 10^4$ \cmt, or with $v_s \sim 15$ \kms\, and
higher pre-shock densities of $n_H = 10^5$ \cmt. Such models, however, still underestimate
the column densities of the near-IR transitions: the post-shocked \h\ gas density  
remains lower than the NIR transitions critical density and the atomic hydrogen
produced from \h\ dissociation is not high enough to populate the vibrational
levels to equilibrium conditions.

The difficulty of finding a suitable single model that reproduce the derived physical 
conditions is likely related to possible geometrical effects and to the fact that multiple shocks with different velocities might be present along the line of sight. It would be indeed  interesting to explore whether bow-shock models might be able to predict the averaged physical characteristics 
along the line of sight that we infer from our analysis.

\subsection{Flow energetics}

\h\ emission represents one of the main contributor to the energy radiated
away in shocks along outflows from very young stars. Kaufman \& Neufeld (1996) predicted that
between 40 and 70\% of the total shock luminosity is emitted in \h\, lines
for shocks with pre-shock density lower than 10$^5$ \cmt\ and shock velocities larger
than 20 \kms, the other main contributions being in CO and H$_2$O rotational emission. 
This has been also observationally tested by Giannini et al. (2001) who measured the 
relative contribution of the different species to the outflow cooling in a sample of class 0 
objects observed with ISO-LWS. 

We will discuss here the role of the \h\, cooling in the global radiated energy of the L1157 outflow.
From the best fit model obtained for the knots A and C, we have derived the total, extinction corrected,
\h\ luminosity by integrating over all the ro-vibrational transitions considered by our
model. $L_{\rm H_2}$ is found
to be 8.4$\times$10$^{-2}$ and 3.7$\times$10$^{-2}$ L$_\odot$ for the A and C knots, respectively. Out of this total
luminosity, the contribution of only the rotational lines is 5.6$\times$10$^{-2}$(A) and 2.7$\times$10$^{-2}$(C) L$_\odot$,
which means that in both cases they represent about 70\% of the total \h\, luminosity.

N09 have found that the total luminosity of the \h\ rotational lines from S(0) to
S(7), integrated over the entire L1157 outflow, 
amount to 0.15 L$_\odot$. If we take into account an additional 30\% of contribution from the v$>$0 
vibrational levels, we estimate a total \h\ luminosity of 0.21 L$_\odot$. This is a 30\% larger
than the total \h\ luminosity estimated by Caratti o Garatti (2006) in this outflow, 
assuming a single component gas at temperature between 2000 and 3000 K that fit the NIR \h\ lines. 

If we separately compute the \h\ luminosity
in the two outflow lobes, we derive $L_{\rm H_2}$ = 8.5$\times$10$^{-2}$ L$_\odot$ in the blue lobe and 1.3$\times$10$^{-1}$ L$_\odot$ in the red
lobe. Comparing these numbers with those derived in the individual A and C knots, 
we note that the A knot alone contributes to most of the \h\ luminosity in the blue lobe. By contrast,
the \h\ luminosity of the red lobe is distributed among several peaks of similar values. 
This might suggest that most of the energy carried out by the blue-shifted jet is
released when the leading bow-shock encounters a density enhancement at the position of the A knot. 
On the other hand, the red-shifted gas flows more freely without large density discontinuities, and 
the corresponding shocks are internal bow-shocks, all with similar luminosities.

The integrated luminosity radiated by CO, H$_2$O and OI in L1157 has been estimated, through ISO and
recent Herschel observations, as $\sim$0.2 L$_\odot$ (Giannini et al. 2001, Nisini et al. 2010), 
which means that \h\ alone contributes about 50\% of the
total luminosity radiated by the outflow. 
Including all contributions, the total shock cooling along the L1157 outflow amounts to about 0.4 L$_\odot$, i.e. $L_{cool}/L_{bol}$ $\sim$ 5$\times$10$^{-2}$, assuming $L_{bol}$=8.4 L$_\odot$ 
for L1157-mm (Froebrich 2005). This ratio is consistent with the range of values derived from other class 0 sources 
from ISO observations (Nisini et al. 2002).

The total kinetic energy of the L1157 molecular outflow 
estimated by Bachiller et al. (2001) amounts to 0.2 L$_\odot$ without any correction for the
outflow inclination angle, or to 1.2 L$_\odot$ if an inclination angle of 80 degrees 
is assumed. Considering that the derivation of the L$_{kin}$ value has normally
an uncertainty of a factor of five (Downes \& Cabrit 2007), we conclude that the mechanical energy flux 
into the shock, estimated as $L_{cool}$,  is comparable to the kinetic energy of the swept-out 
outflow and thus that the shocks giving rise to the H$_2$ emission have 
enough power to accelerate the molecular outflow.

The total shock cooling derived above can be also used to infer the momentum flux 
through the shock, i.e. $\dot{P}$ = 2$L_{cool}$/V$_s$, where V$_s$ is the shock velocity that
we can assume, on the basis of the discussion in the previous section, to be of the order of 20 \kms.
Computing the momentum flux separately for the blue and red outflow lobe, we derive 
$\dot{P}_{red} \sim$ 1.7$\times$10$^{-4}$ and $\dot{P}_{blue} \sim$ 1.1$\times$10$^{-4} $M$_\odot$ yr$^{-1}$ \kms. 
In this calculation, we have assumed that the contribution from cooling species
 different from \h, as estimated by ISO and Herschel, is distributed among the two lobes 
 in proportion to the \h\ luminosity. If we assume that the molecular 
 outflow is accelerated at the shock front through momentum conservation, then the
 above derived momentum flux should results comparable to the thrust of the outflow,
 derived from the mass, velocity and age measured through CO observations.
The momentum flux measured in this way by Bachiller et al. (2001) is 1.1$\times$10$^{-4}$and 2$\times$10$^{-4}$ M$_\odot$ yr$^{-1}$ \kms
in the blue and red lobes, respectively, i.e. comparable to our derived values. It is interesting to note that 
the $\dot{P}$ determination from the shock luminosity confirms the asymmetry between
the momentum fluxes derived in two lobes. As shown by Bachiller et al. (2001), the L1157 red lobe 
has a 30\% smaller mass with respect to the blue lobe, but a higher momentum flux due to the larger flow velocity. The northern red lobe is in fact more extended than the southern lobe:
however, given the higher velocity of the red-shifted gas, the mean kinematical ages of the two lobes is very
similar.

\section{Conclusions \label{conclusions}}

We have analysed the \h\ pure rotational line emission, from S(0) to S(7),
along the outflow driven by the L1157-mm protostar, mapped with the Spitzer - IRS 
instrument. The data have been analysed assuming a gas temperature stratification where the \h\ column 
density varies as $T^{-\beta}$ and 2D maps of the \h\ column density,
ortho-to-para ratio (OPR) and temperature spectral index $\beta$
have been constructed. 
Further constraints on the physical conditions of the shocked gas have been derived 
in two bright emission knots by combining the Spitzer observations with near-IR 
data of \h\ ro-vibrational emission. Finally, the global \h\ radiated energy of the
outflow has been discussed in comparison with the energy budget of the associated
CO outflow.

The main conclusions derived by our analysis are the following:
\begin{itemize}
\item \h\ transitions with $J_{lower} \le$ 2 follows the morphology of the CO molecular
outflow, with peaks correlated with individual CO clumps and more diffuse
emission that delineates the CO cavities created by the precessing jet. 
Lines with higher $J$ are localized on the shocked peaks, presenting a morphology
similar to that of the \h\, 2.12\um\, ro-vibrational emission.
\item  Significant variations of the derived parameters are observed along the flow. 
The \h\ column density ranges between 5$\times$10$^{19} $ and 3$\times$10$^{20} $\cmd: 
the highest values are found in the blue-shifted lobe, suggesting that here the outflowing
gas is compressed due to the impact with a high density region.
Gas components in a wide range  of temperature values, from $ \sim $ 250 to $ \sim $ 1500 K
contribute to the \h\ emission along individual lines of sight. The largest range
of temperature variations is derived towards the intensity peaks closer to the 
driving source, while a more uniform temperature distribution, with $ T $
between $ 400 $ and  $ 1000$ K, is found at the tip of the northern outflow lobe.
\item The OPR is in general lower than the equilibrium value at high temperatures and spans a range from $\sim$0.6 to 2.8, with the lowest values
found in low temperature plateau regions between consecutive intensity peaks. 
As in previous studies, we also found the presence of regions at low OPR (1.5-1.8) 
but with relatively high temperatures. These might represent zones subject to recent shocks 
where the OPR has not had time yet to reach the equilibrium value.

\item Additional shock parameters have been derived in the two bright near-IR 
knots A and C,  located 
in the blue- and red-shifted outflow lobes, where the mid- and near-IR \h\ 
data have been combined. 
The ratio between mid- and near-IR lines is very sensitive to the molecular  plus atomic hydrogen particle density. A high 
abundance of atomic hydrogen (H/\h $\sim$ 0.1-0.3) is implied by the 
the observed \h\ column densities if we assume n(\h) values as derived by independent 
mm observations. With this assumption, the cooling lengths  of the shock result 
of the order of 7$\times$10$ ^{14} $  and 10$ ^{15} $ cm for the A and C knot, respectively.
The distribution of OPR values as a function of temperature  and the 
derived abundance of atomic hydrogen, implies that the shock passing time is of the
order of 100 yr for knot A and 1000 yr  for  knot C, given the assumption that the para-to-ortho
conversion occurs through reactive collisions with atomic hydrogen.
We find that planar shock models, either of C- or J-type, are
not able to consistently reproduce all the physical parameters derived from our analysis 
of the \h\ emission. 
\item Globally, \h\ emission contributes to about 50\% of the total shock radiated energy in the L1157 outflow. We find that the momentum flux through the shocks derived from the radiated luminosity is
comparable to the thrust of the associated molecular outflow, supporting a scenario
where the working surface of the shocks drives the molecular outflow. 
\end{itemize}

\acknowledgments

This work is based on observations made with the Spitzer Space Telescope, which is operated by the Jet Propulsion Laboratory, California Institute of Technology under a contract with NASA. Financial support from contract ASI I/016/07/0 is acknowledged. 

\bibliographystyle{plainnat}

%
\clearpage
%
%
%
%
%
%

\clearpage

\begin{deluxetable}{cccccccc}
\tablecaption{\h\, line brightness in selected positions \label{fluxes}}
\tablewidth{0pt}
\tablehead { Transition & $\lambda$($\mu$m)  &Module  & \multicolumn{5}{c}{Intensity$^a$ (10$^{-5}$erg\,cm$^{-2}$\,s$^{-1}$\,sr$^{-1}$)}\\
\cline{4-8}
& & & (+20,$-$56) & ($-$2,+30) & ($-$32,+58) & ($-$26,+114)& ($-$30,+140)$^b$
 }
\startdata
 S(0)\,$J=2-0$   & 28.21   & LH  &0.70 & 0.09 & 0.34 & 0.40 & 0.29\\
 S(1)\,$J=3-1$   & 17.03 &  SH    &4.47 & 0.43 & 3.23& 3.50 & 4.30 \\
 S(2)\,$J=4-2$   & 12.28  &  SH   & 8.70 & 0.30  & 4.05& 6.63 &7.95 \\
 S(2)\,$J=4-2$    & 12.28  &  SL   & 7.05 & 0.34 & 3.58 & 4.39 &7.11 \\
 S(3)\,$J=5-3$    & 9.66  &  SL    &21.85 & 1.82 &14.61& 9.38 & 17.86 \\
 S(4)\,$J=6-4$    &  8.02 &  SL   &12.89 & $<$ 2.1$^c$  &5.38 & 5.47 & 8.30 \\
 S(5)\,$J=7-5$    & 6.91 &  SL    & 27.34 &1.76  &11.91& 6.72 &12.74 \\
 S(6)\,$J=8-6$    & 6.11 &  SL    &5.74 &$<$  1.3$^c$ &3.00 & 3.32 & 3.30 \\
 S(7)\,$J=9-7$     & 5.51  &  SL   &18.11 & 0.97  &6.63& 3.35 &3.10 \\

\enddata
\tablenotetext{a}{Intensity averaged within a 20$\arcsec$ FWHM Gaussian aperture around the considered position. Absolute errors are of the order of 20\%.}
\tablenotetext{b}{Offsets in arcsec with respect to the L1157-mm source.
These positions are associated to individual near-IR/mm emission clumps, as indicated
in Table 2.}

\tablenotetext{c}{2$\sigma$ upper limit}
\end{deluxetable}

\begin{deluxetable}{ccccccccc}
\tablecaption{Parameters derived from rotational diagrams$^a$ \label{param}}
\tablewidth{0pt}
\tablehead{Offset & N(H$_2$) & OPR & $T_{cold}^b$ & $T_{warm}^b$ & $T_{med}^b$ & $\beta$ & Association$^c$\\
(arcsec) & (cm$^{-2}$)  &  &( K) &(K) &(K) & \\
}
\startdata
%
(+20,$-$56) & 2.5$\times$10$^{20}$ & 1.8 & 300 & 1400 & 850 & 4.0 & mm-B1; NIR-A\\
($-$2,+30) & 5.0$\times$10$^{19}$ &  2.4 & 246& 1370 & 970 & 3.9&\\
($-$32,+58) & 1.0$\times$10$^{20}$ & 2.2 &340 & 1300 & 870 &4.2 & mm-R0; NIR-C\\
($-$26,+114) & 1.0$\times$10$^{20}$ & 1.4 & 370  & 1100 & 735 & 5.0& mm-R1\\
($-$30,+140) & 8.0$\times$10$^{19}$ & 1.6 & 410 & 1049 & 715& 5.4 & mm-R; NIR-D\\

\enddata
\tablenotetext{a}{Typical uncertainty in the parameters is as follows: N(H$_2$) $\pm$ 0.1 dex, 
OPR $\pm$ 0.2, $T \pm$ 50 K, $\beta \pm 0.2$.}
\tablenotetext{b}{$T_{cold}$ and $T_{warm}$ are the temperatures derived from linearly fitting 
the S(0)-S(1)-S(2) and S(5)-S(6)-S(7) lines. $T_{med}$ is
the average temperature derived from a liner fit through all the \h\ lines. }
\tablenotetext{c}{Association with mm clumps and with NIR-knots, as defined in Bachiller et al. (2001)
and Davis \& Eisl\"offel (1995), respectively.}
\end{deluxetable}

\begin{deluxetable}{lcccc}
\tablecaption{Inferred shock parameters$^a$ \label{shock}}
\tablewidth{0pt}
\tablehead{Parameter & \multicolumn{2}{c}{knot A} & \multicolumn{2}{c}{knot C} \\
\cline{2-5} 
 & A & B & A & B
}

\startdata
$n(H_2)$ (\cmt) & 3$\times$10$^5$ & 10$^7$& 10$^5$ & 2$\times$10$^6$\\
$H/H_2$ & 0.3 & ... & 0.1 & ...\\
$OPR_i$ & \multicolumn{2}{c}{0.6} & \multicolumn{2}{c}{0.6} \\
$L$ (cm)&  7$\times$10$^{14}$ & 10$^{13}$& 10$^{15}$ & 5$\times$10$^{13}$\\
$\tau$ (yr) & $\sim$10$^{2}$ & $\sim$10$^{3}$ & 10$^{3}$ & $\sim$2$\times$10$^4$\\
$OPR_{med}$ & \multicolumn{2}{c}{1.8} & \multicolumn{2}{c}{2.2} \\
$T_{med}$ (K) &\multicolumn{2}{c}{850} & \multicolumn{2}{c}{870} \\
\enddata
\tablenotetext{a}{See text for details on the various parameters.}
\tablenotetext{A}{Model that assumes the \h\ particle density from Nisini et al. (2007) and 
a variable H/\h\ ratio. }
\tablenotetext{B}{Model that assumes negligible hydrogen in atomic form.}
\end{deluxetable}

\begin{figure}
  \centering \includegraphics[scale=0.7,angle=-90]{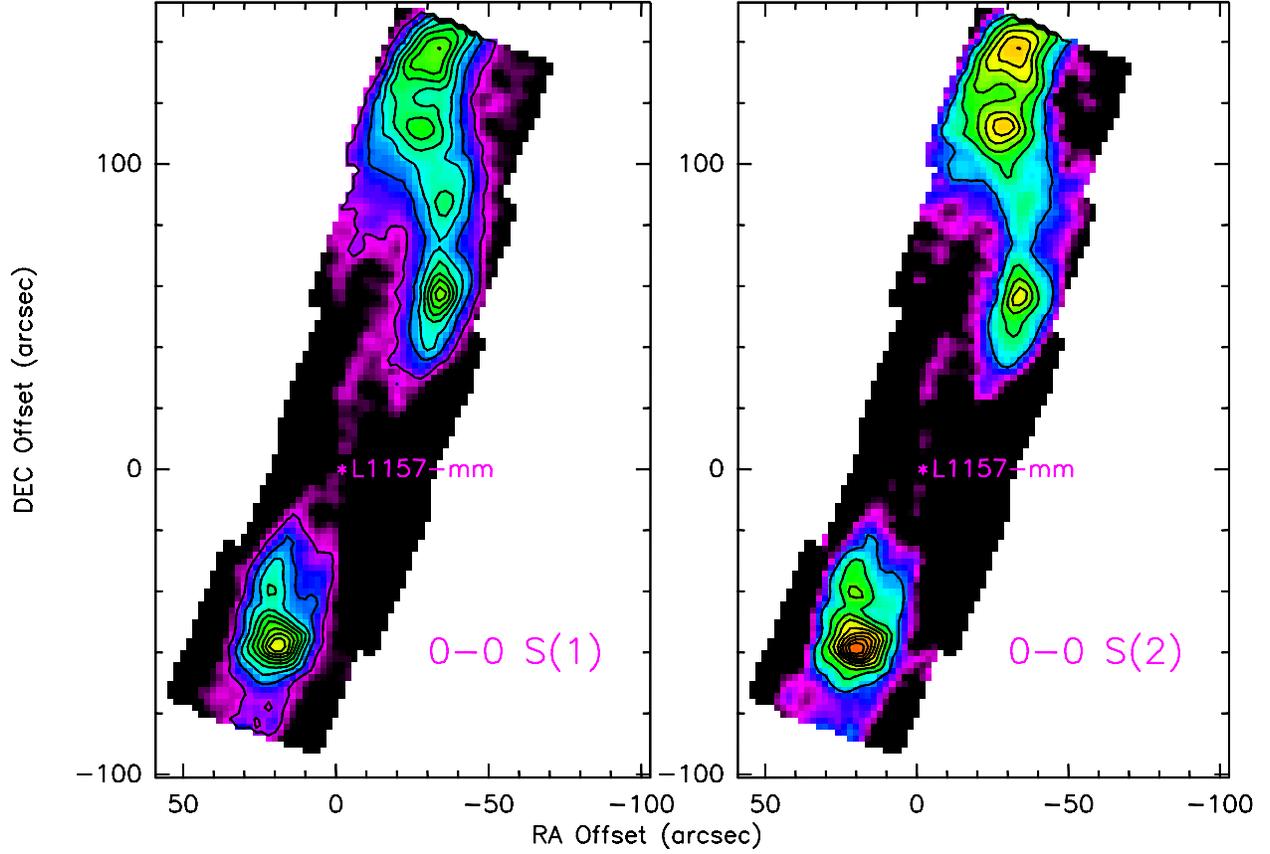}
  \caption{L1157 maps of the 0--0 S(1)17.035 \um\  and S(2)12.29 \um\ lines observed with the Spitzer IRS-SH module. 
In both images contours are displayed in steps of 10\% of the peak intensity value, which is 8.2$\times$10$^{-5}$ and 1.9$\times$10$^{-4}$ erg\,s$^{-1}$cm$^{-2}$sr$^{-1}$ for the S(1) and S(2) maps, respectively. The average rms noise in the two images is of the order of 1.5$\times$10$^{-6}$ erg\,s$^{-1}$cm$^{-2}$sr$^{-1}$.
  \label{fig:h2image}
  }

\end{figure}
\begin{figure}
  \centering \includegraphics[scale=0.6,angle=-90]{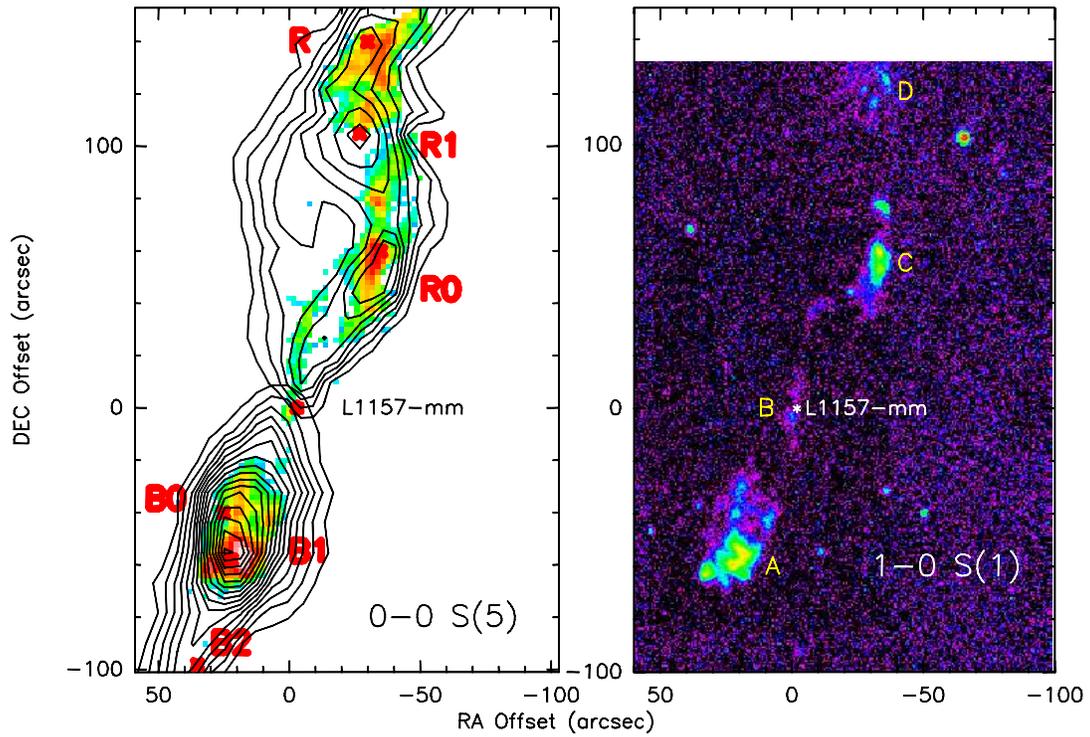}
  \caption{(left) Map of the 1--0 S(5) line at 6.9 \um\ with superposed contours of the integrated CO (2-1) emission (from Bachiller et
  al. 2001). The position of the molecular condensations (bullets) identified in (Bachiller et al. 2001) are labelled. (right) Map of the 1-0 S(1) line at 2.12 \um\ (Caratti o Garatti et al. 2006).The main group of NIR knots, labelled from A to D according to 
Davis \& Eisl\"offel (1995) , are indicated.
  \label{fig:h2co}
  }
\end{figure}

\begin{figure}
  \centering \includegraphics[scale=1,angle=-90]{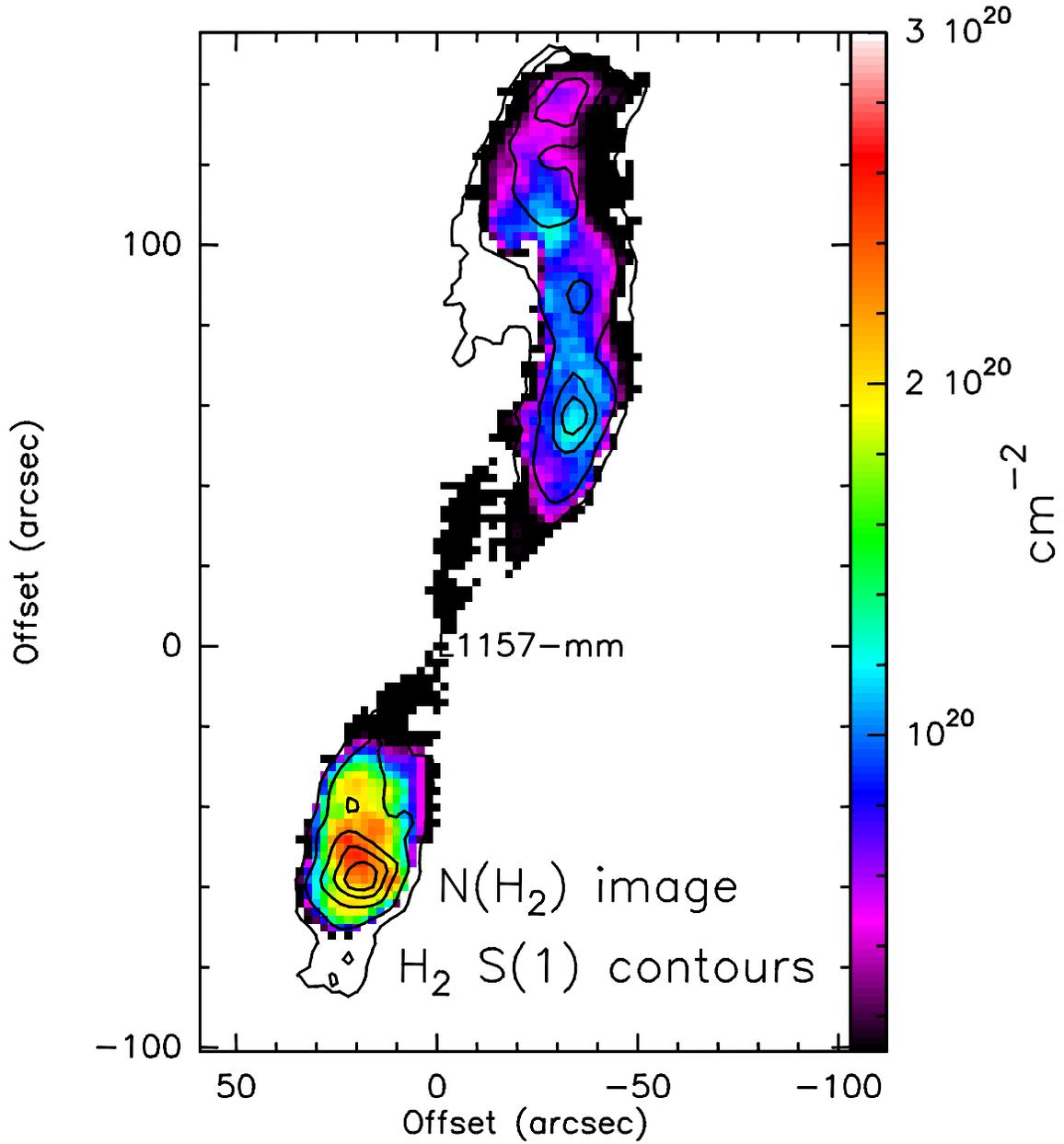}
  
  \caption{Map of the total H$_2$ column density. Values have been derived only in the pixels where at least four transitions with an  S/N larger than 3 have been detected.
Contours refer to the intensity of the S(1) line.
  \label{fig:ncol}
  }
\end{figure}

\begin{figure}
  \centering \includegraphics[scale=1,angle=-90]{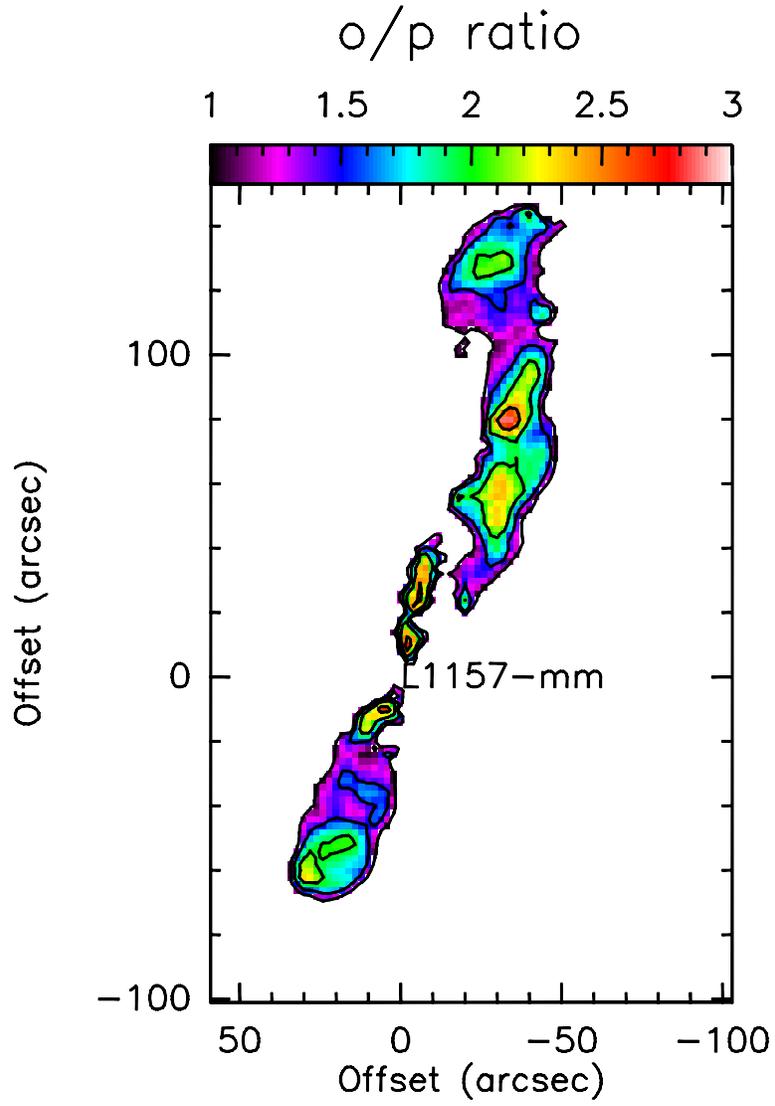}
  \caption{As in Fig. 3 for the map of the ortho-to-para ratio. 
  \label{fig:b_tmin}
  }
\end{figure}

\begin{figure}
  \centering \includegraphics[scale=0.9,angle=0]{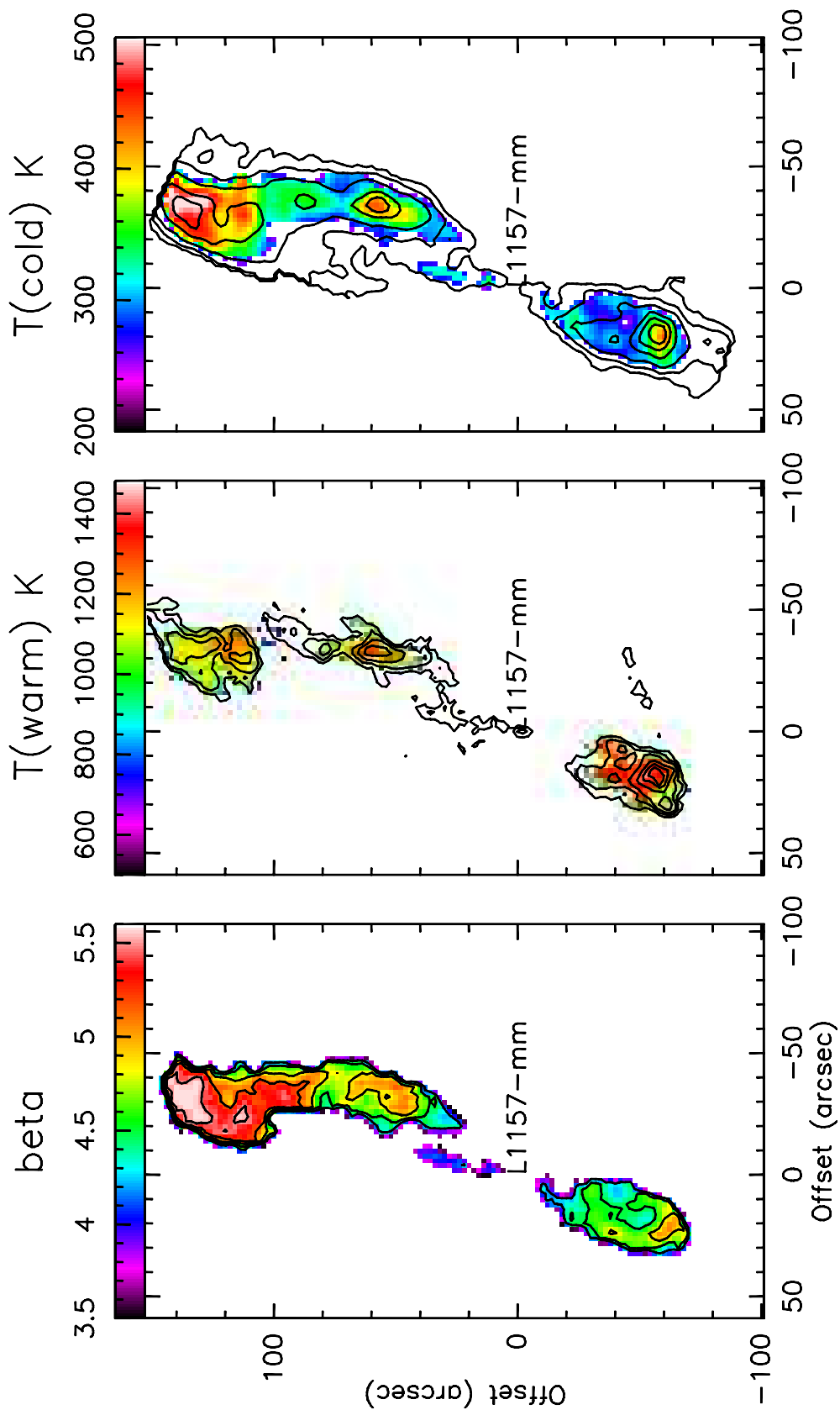}
  \caption{Maps of the fitted temperature power law spectral index $\beta$ (a), and of the "cold" and "warm"
temperature components, derived from a linear fit through the OPR corrected S(0)/S(1)/S(2) lines and S(5)/S(6)/S(7) lines, respectively. Contours of the intensity of the S(5) and S(1) lines 
are superimposed on the T$_{warm}$ and T$_{cold}$ images, respectively.
  \label{fig:op_tcold}
  }
\end{figure}


\begin{figure}
  \centering \includegraphics[scale=1,angle=0]{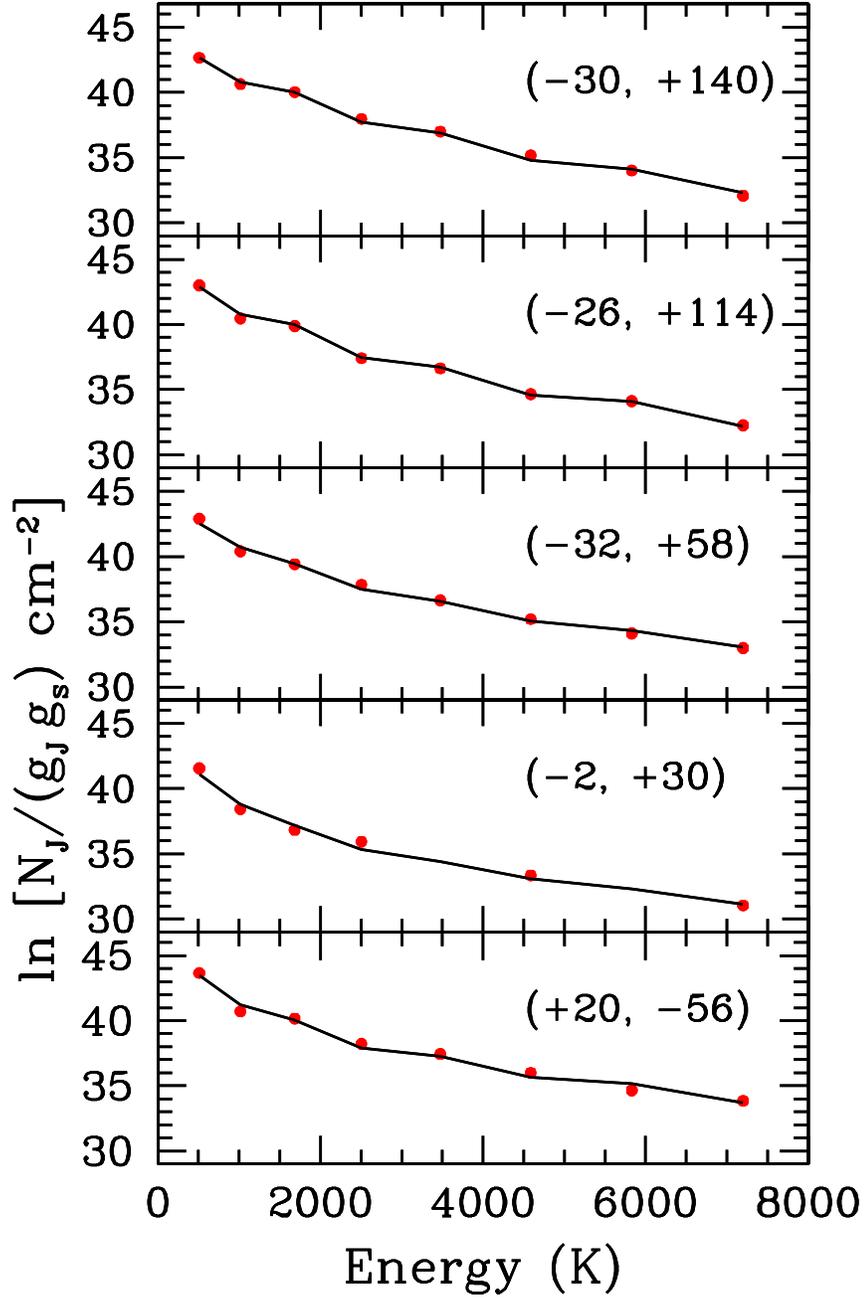}
  \caption{H$_2$ rotational diagrams obtained at the offset positions indicated in each panel. The filled circles correspond to the observations while the solid line is the best LTE fit obtained in each position. The parameters of each fit are summarized in Tab. \ref{param}
  \label{fig:fit_nir}
}
\end{figure}

\begin{figure}
  \centering \includegraphics[scale=0.8,angle=0]{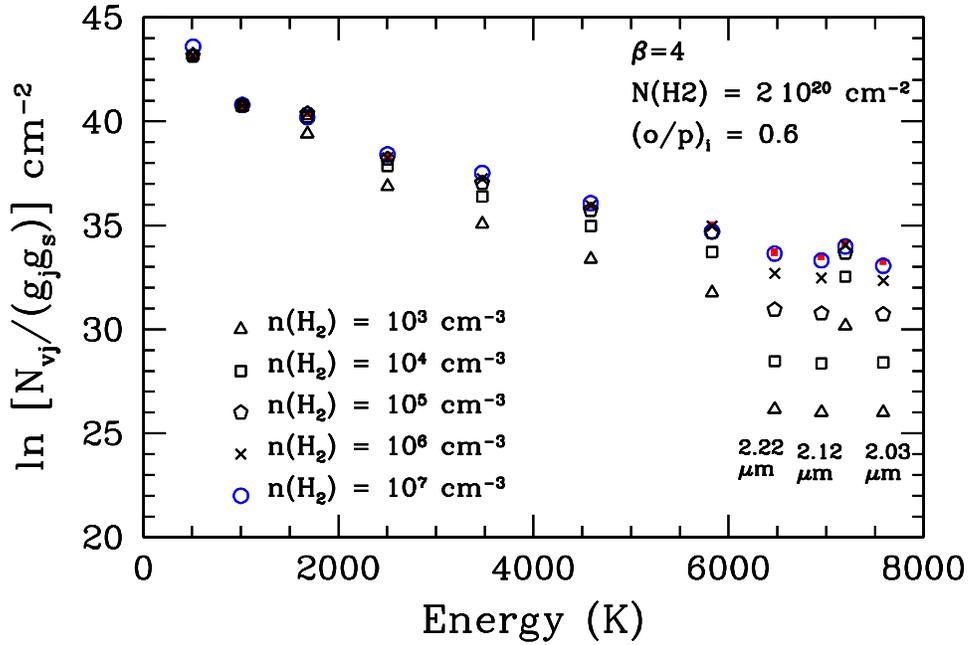}
  \caption{H$_2$ rotational diagram showing the results of the NLTE model at different H$_2$ densities for the Spitzer observed rotational lines and three bright NIR lines at excitation temperatures between 6000 and 8000 K. The model assumes a slab where the temperature decreases as a power low with index $\beta$ and where the o/p 
ratio varies as a function of temperature. The gas is assumed fully molecular (H/H$_2$=0). Filled squares correspond to the observations in a 20$\arcsec$ aperture centered towards the knot A in the south-east lobe (see Fig.\ref{fig:h2image}). Different symbols indicate the model results at different densities: the other parameters are indicated in the figure and are kept fixed. It should be noted how the near-IR lines are much more sensitive to density
variations than the pure rotational lines, due to their different critical densities.
  \label{fig:plot_dens}
  }
\end{figure}

\begin{figure}
  \centering \includegraphics[scale=0.8,angle=0]{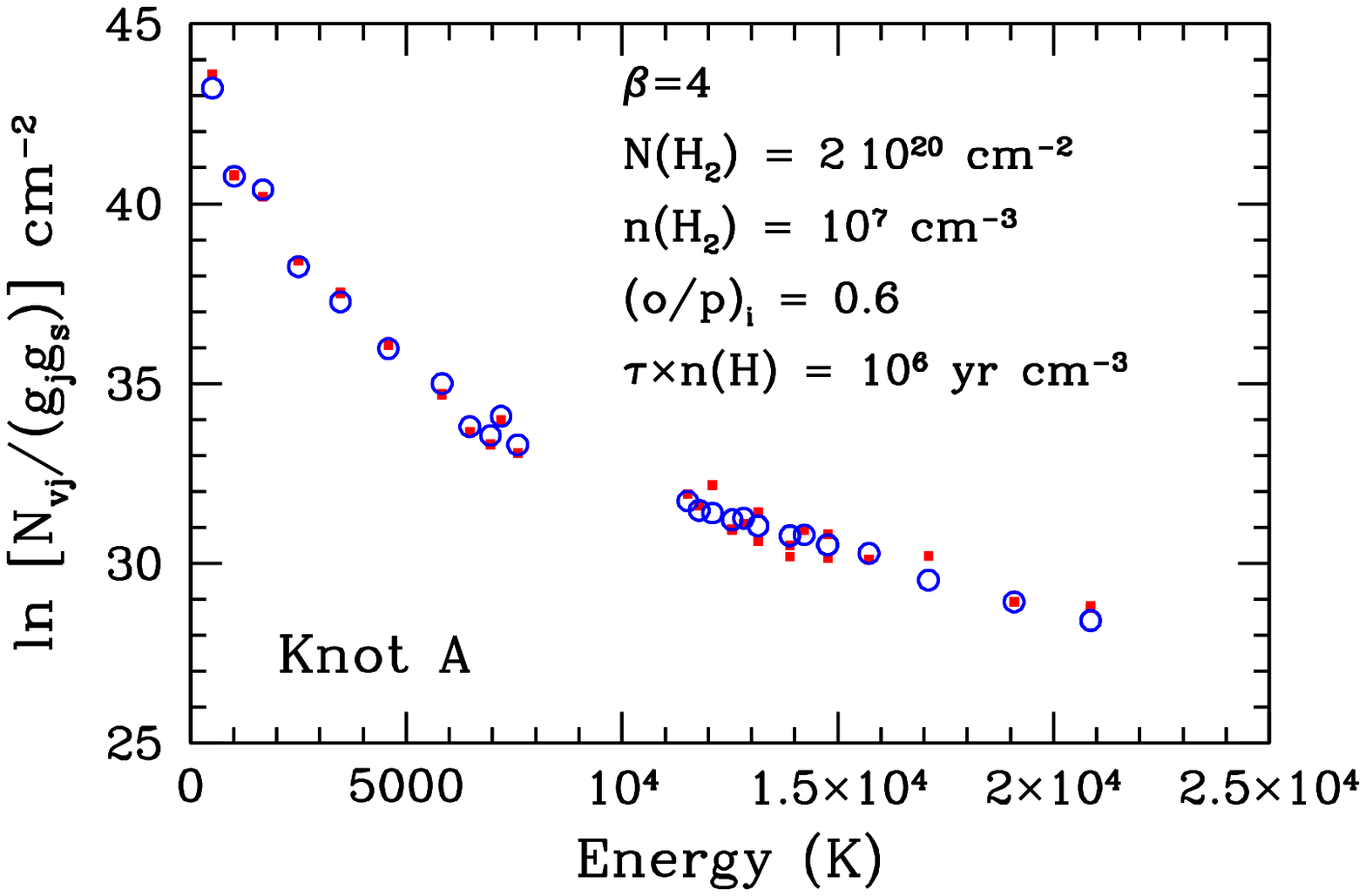}
  \centering \includegraphics[scale=0.8,angle=0]{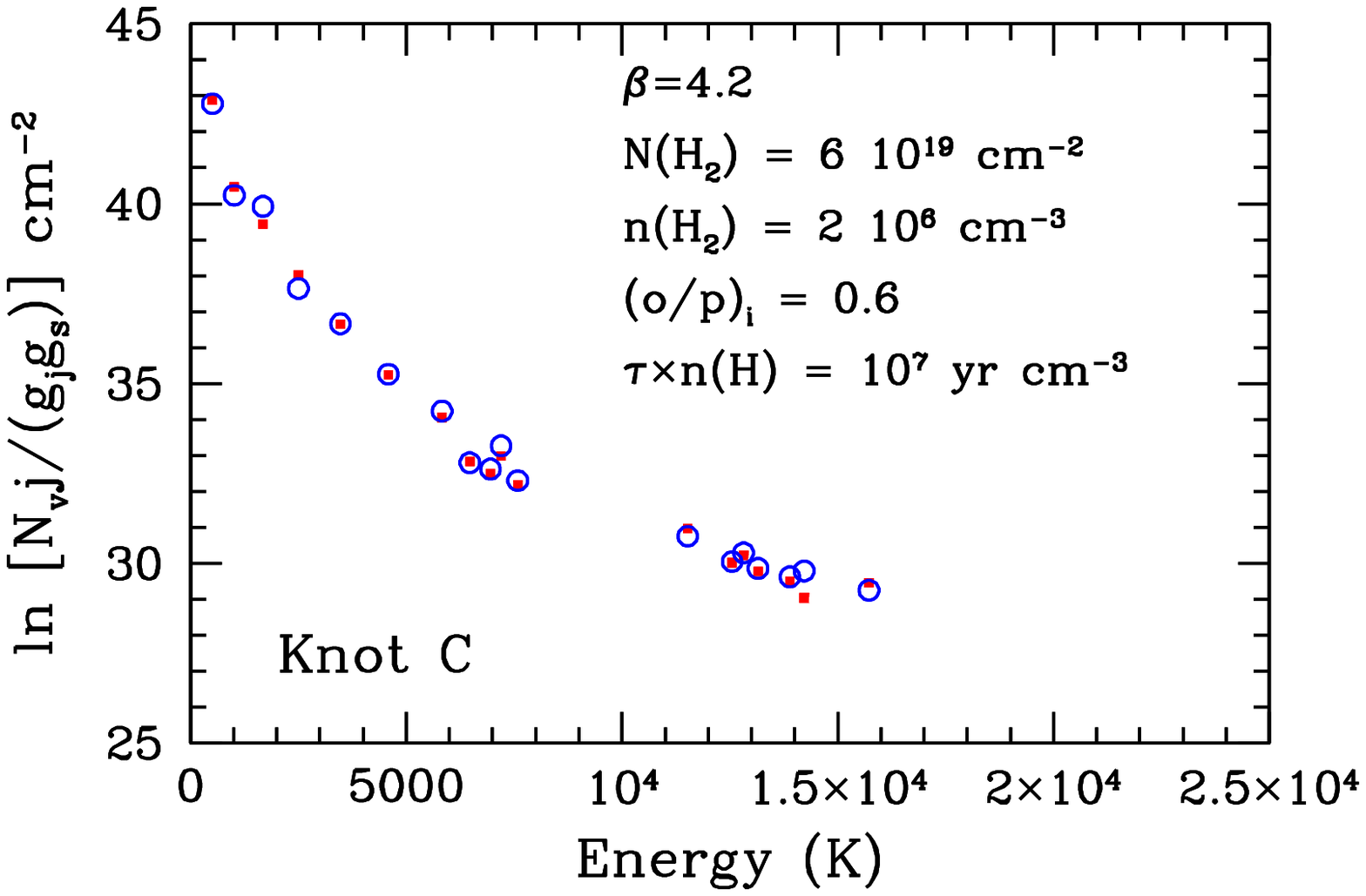}
  \caption{H$_2$ rotational diagrams showing the best model fits (open circles) through both the Spitzer and near-IR data (filled squares) in 20$\arcsec$ apertures centered towards the knots A and C (see Fig.\ref{fig:h2image}). The gas is assumed fully molecular (H/H$_2$=0).
  \label{fig:ACfit}
  }
\end{figure}

\begin{figure}
  \centering \includegraphics[scale=1,angle=0]{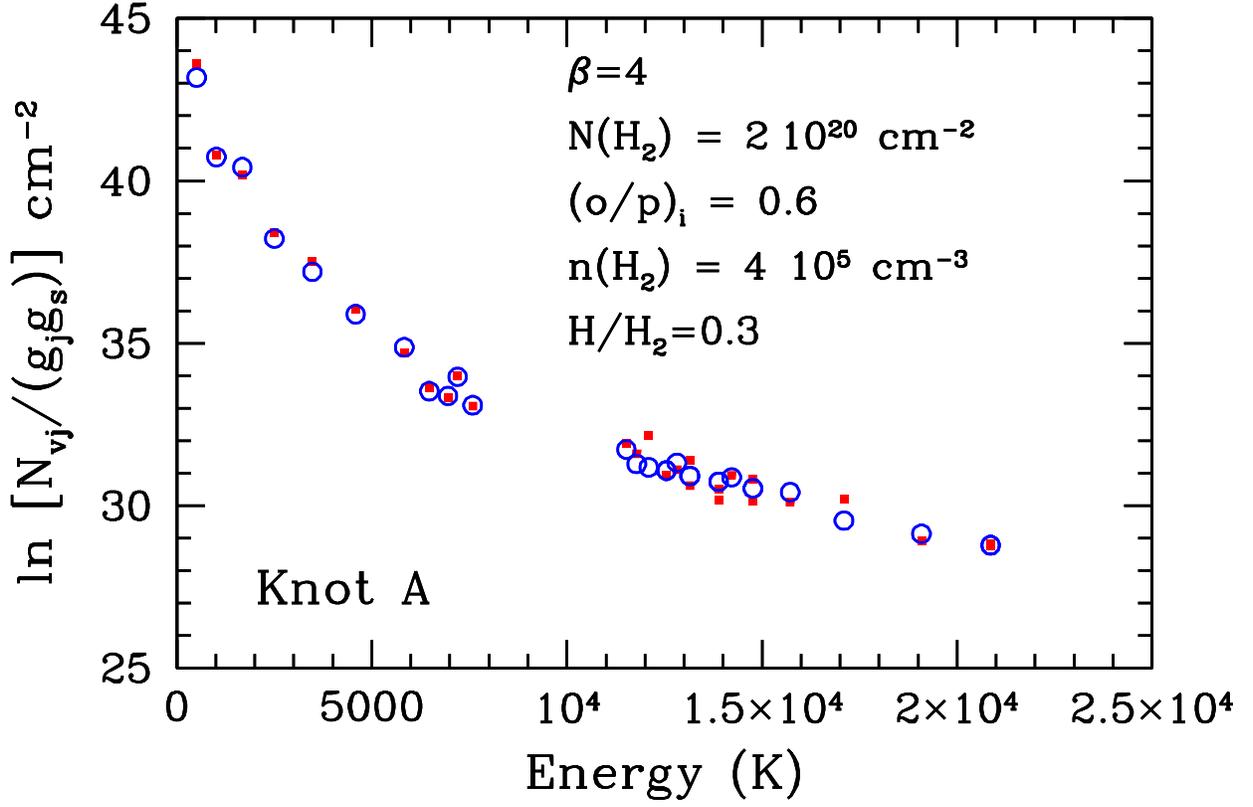}
  \caption{As in Fig. \ref{fig:ACfit} but keeping the H$_2$ density equal to the value of 4$\times$10$^5$ \cmt\, as 
measured from SiO (Nisini et al. 2007), and varying the H/H$_2$ ratio. Agreement with
the data is obtained with a neutral hydrogen abundance $\sim$ 30\% of the H$_2$ density.
  \label{fig:average_opr_trot}
  }
\end{figure}
\end{document}